\documentclass[amsmath,amssymb,nofootinbib]{revtex4}

\usepackage{latexsym,comment,verbatim}
\usepackage{amssymb}
\usepackage{amsfonts}
\usepackage{amsmath,color}
\usepackage[dvips]{graphicx}
\usepackage{bm}

\def\mb#1{\mathbf{#1}}

\def\ber{\begin{eqnarray}}
\def\eer{\end{eqnarray}}
\def\beq{\begin{equation}}
\def\eeq{\end{equation}}

\def\ed{\end{document}}
\def\di#1#2{\frac{\mathrm{d} #1}{\mathrm{d}#2}}
\begin{document}

\title{A Note on the Sagnac Effect for Matter Beams}

\author{Matteo Luca Ruggiero}
\email{matteo.ruggiero@polito.it}
 \affiliation{DISAT, Politecnico di Torino, Corso Duca degli Abruzzi 24, Torino, Italy\\
 INFN, Sezione di Torino, Via Pietro Giuria 1, Torino, Italy}
\author{Angelo Tartaglia}
\email{angelo.tartaglia@polito.it}
 \affiliation{DISAT, Politecnico di Torino, Corso Duca degli Abruzzi 24, Torino, Italy\\
 INFN, Sezione di Torino, Via Pietro Giuria 1, Torino, Italy}

\date{\today}

\begin{abstract}
We study the Sagnac effect for matter beams, in order to estimate the kinematic corrections to the basic formula, deriving from the position and the extent of the interferometer, and discuss the analogy with the Aharonov-Bohm effect. We show that the formula for the Sagnac time delay is the same for matter and light beams in arbitrary stationary space-times, provided that a suitable condition on the speed of the beams is fulfilled. Hence, the same results obtained for light beams apply to matter beams.
\end{abstract}

\maketitle

%------------------------Section-------------------------
\section{Introduction}\label{sec:sdec}
%------------------------Section-------------------------

In a recent paper \cite{sagnac14} we studied the Sagnac effect \cite{sagnac05,sagnac13} for light rays, in order to  evaluate  if the higher order relativistic corrections of kinematic origin could be relevant for actual terrestrial experiments (e.g. \cite{ginger11,ginger12}): in particular, we focused on the role of the  position and extent of the interferometer in the rotating frame, and discussed the analogy with the Aharonov-Bohm effect. We showed  that the analogy with the Aharonov-Bohm effect is true to lowest approximation order only, and that the influence of  position and extent of the interferometer is  negligible for current experiments. So, for actual experiments, the Sagnac time delay, i.e. the difference between the propagation times of the co-rotating and counter-rotating beam, as measured by an observer at rest in the rotating frame,  can be safely written as
\begin{equation}
\Delta t=4\frac{\mathbf{\bm{\Omega} \cdot
S}}{c^{2}} \label{eq:sagnac2}
\end{equation}
in terms of  $\bm{\Omega }$, the (constant) rotation rate of the interferometer with respect to an inertial  frame, and  $\mathbf{S}$, the vector associated to the area enclosed by the light path.

In this paper we want to show that the same results hold true for matter beams, i.e. for (time-like) particles propagating along opposite directions in a uniformly rotating interferometer. The key point is the validity of the Sagnac time delay formula independently of the physical nature of the interfering beams: in previous works (see  \cite{rizzi03b,RRinRRF} and references therein), one of us discussed this issue (the so called ``universality'' of the Sagnac effect) in the case of matter and light beams propagating in a circular interferometer, in flat space-time.  A more general approach, that can be used also in curved space-time, was carried out in \cite{scorgie}. Here, we first rephrase the latter approach in order to show that, in arbitrary stationary space-time, the Sagnac effect does not depend on the physical nature of the propagating beams, provided that suitable kinematic conditions are fulfilled, and then we generalize the results of the previous paper \cite{sagnac14} to matter beams.

%------------------------Section-------------------------
\section{Propagation Times and the Sagnac Effect}\label{sec:prop}
%------------------------Section-------------------------

The physical situation that we are going to consider is the following one. An interferometer is at rest in a reference frame (the \textit{interferometer frame}) and it simultaneously emits two beams:  they propagate in opposite directions along the same path and reach the emission point at different times: we  call \textit{Sagnac time delay} the proper time difference between the two times of arrival, measured in the interferometer frame.

Actually, the interferometer frame could be just a rotating frame, such as a turntable in the laboratory, or a more general frame, such as a terrestrial laboratory, where the rotation effects have both kinematic and gravitational origin.  In any case,   in the interferometer frame we choose a set of adapted   coordinates  $\{x^{\mu}\}=\{x^{0},x^{i}\}=\{ct,\mb x\}$ and we write the squared line-element in the form\footnote{We use the following notation: Greek (running from 0 to 3) and Latin (running from 1 to 3) indices
denote space-time and spatial components, respectively;
letters in boldface like $\mb x$ indicate spatial vectors.}
\beq
ds^{2}=g_{00}c^{2}dt^{2}+2g_{0i}cdtdx^{i}+g_{ij}dx^{i}dx^{j} \label{eq:metricastazionaria}
\eeq
The space-time metric does not depend on time because of stationarity, so that $g_{\mu\nu}=g_{\mu\nu}(\mb x)$; we choose the signature $(-,+,+,+)$, so that $g_{00}<0$.  The above metric is not \textit{time-orthogonal}, because $g_{0i} \neq 0$. Actually, the off-diagonal terms $g_{0i}$ depend both on the global configuration of the space-time and on the rotational features of the reference frame: in the case of a rotating frame in flat space-time, they depend on the rotation rate; more in general, they express the rotation rate of the frame with respect to a Fermi-Walker tetrad (see e.g. \cite{rindlerperlick}). As a consequence, we can speak of Sagnac effect for a space-time in the form (\ref{eq:metricastazionaria}) because of its rotational features.

\begin{comment}
We use this notation:  in the 3-dimensional space  of the metric (\ref{eq:metricastazionaria}), the length element is
\beq
d\ell^{2}=g_{ij}dx^{i}dx^{j}  \label{eq:dl2}
\eeq
Then, the  particles have unit tangent vectors
\beq
\ell^{i}=\di{x^{i}}{\ell}, \label{eq:defell1}
\eeq
and we may write the components of the coordinate speed in the form
\beq
u^{i} = \di{x^{i}}{t} = u \ell^{i} \label{eq:defiu1}
\eeq
with
\beq
u^{2}={u^{i}u_{i}=g_{ij}u^{i}u^{j}}=\di{\ell^{2}}{t^{2}} \label{eq:defiu2}
\eeq

Let $d\tau$ be the proper time interval, measured by the particles along the path; of course, $d\tau$ is zero for light beams. On substituting this expression in the line-element (\ref{eq:metricastazionaria}), taking into account (\ref{eq:defell1}), we obtain

\beq
-c^{2} d\tau^{2}=ds^{2}= g_{00}c^{2}dt^{2}+2g_{0i}\ell^{i}cdtd\ell+d\ell^{2}
\label{eq:metricaell1}
\eeq

\end{comment}

In what follows, we study time-like (for matter beams) and light-like (for light beams) particles, moving in the space-time metric (\ref{eq:metricastazionaria}), to calculate the time elapsed for a complete round trip in the interferometer path. To this end, we have to impose some condition to say that the particles propagating in the two opposite directions are identical but differ only for the direction of propagation: this is naively related to their speed. However, the coordinate speed $\displaystyle w^{i}=\di{x^{i}}{t}$ has not a direct physical meaning. If we want to give an operational meaning (i.e. in terms of observable quantities) to the speed of a particle, we may proceed as follows. Let us consider the coordinate point of the interferometer frame, occupied by the particle at a given time; we introduce an inertial frame, relative to which the point {is at rest:} this is the so-called \textit{Locally Co-Moving Inertial Frame} (LCIF).  In this frame, the proper element of distance  $d\sigma$ and time $dT$ can be defined in terms of the metric elements and coordinates intervals in the interferometer frame by (see \cite{LL, RRinRRF})
\beq
d\sigma^{}=\sqrt{\gamma_{ij}dx^{i}dx^{j}},\quad dT^{}=-\frac 1 c \frac{g_{\mu 0}}{\sqrt{-g_{00}}}dx^{\mu}
\eeq
where $\displaystyle \gamma_{ij}=\left(g_{ij}-\frac{g_{i0}g_{j0}}{g_{00}} \right)$. Indeed, on using these expressions, the line-element (\ref{eq:metricastazionaria}) is locally Minkowskian in the form:
\beq
ds^{2}=d\sigma^{2}-c^{2}dT^{2} \label{eq:metricadsigmadT}
\eeq

In the LCIF an observer  attributes to a particle a speed of magnitude  $v=\di{\sigma}{T}$, i.e. the ratio between the proper element of distance $d\sigma$, traveled in a proper time interval $dT$, and $dT$. In doing so, we have been able to introduce the particle speed $v$, which has a well defined operational meaning and which will be important to fix a natural condition on the properties of the two counter propagating beams, as we are going to discuss.

On substituting in (\ref{eq:metricadsigmadT}), we get
\beq
ds^{2}= \left (1-\frac{c^{2}}{v^{2}} \right) d\sigma^{2} =  \left (1-\frac{c^{2}}{v^{2}} \right)\gamma_{ij} dx^{i}dx^{j}
\eeq
and from  (\ref{eq:metricastazionaria}) we obtain
\beq
 \left (1-\frac{c^{2}}{v^{2}} \right)  \gamma_{ij}dx^{i}dx^{j}= g_{00}c^{2}dt^{2}+2g_{0i}c dt dx^{i}+g_{ij}dx^{i}dx^{j}
 \label{eq:dt1}
\eeq

Eq. (\ref{eq:dt1}) can be solved for the coordinate  time interval $dt$; to this end, we introduce $\displaystyle \beta \doteq v/c$. Notice that for light-like particles, on setting $ds^{2}=0$, we get  $\beta=1$, in agreement with the second postulate of special relativity,  and the left hand side of Equation (\ref{eq:dt1}) is equal to zero.  On using the definition of $\gamma_{ij}$, Eq. (\ref{eq:dt1}) now reads
\beq
0 = g_{00}c^{2}dt^{2}+2g_{0i}c dt dx^{i}+ \left(\frac{1}{\beta^{2}} \gamma_{ij}+\frac{g_{i0}g_{j0}}{g_{00}}\right)dx^{i}dx^{j}
 \label{eq:dt11}
\eeq
from which we obtain the two solutions
\beq
dt_{\pm}= \frac{1}{|g_{00}|c} \left(g_{0i}dx^{i} \pm \frac{1}{\beta}\sqrt{|g_{00}|\gamma_{ij}dx^{i}dx^{j}} \right) \label{eq:dtsol1}
\eeq
Once that the propagation path is known, (\ref{eq:dtsol1}) can be integrated  to obtain the coordinate time interval. Remember that we are interested in the future oriented branch of the light cone: hence we obtain two solutions,  corresponding to the propagation times along opposite directions in the path.  We notice that the properties of the propagating particles (that depend on their physical nature)  appear in (\ref{eq:dtsol1}) through the coefficient $1/\beta$. The coordinate time intervals for the propagation in two opposite directions in the \textit{same}  path $\ell$ can be written as
\beq
t_{+} = \oint_{\ell} dt_{+}, \quad t_{-} =-\oint_{\ell} dt_{-} \label{eq:dtpdtm}
\eeq
So, the difference between the co-rotating ($t_{+}$) and counter-rotating ($t_{-}$) propagation times turns out to be
\beq
\Delta t = t_{+}-t_{-}= \oint_{\ell} \left(dt_{+}+dt_{-} \right) \label{eq:deltapm11}
\eeq
This expression simplifies if we assume that \textit{ the speed $v$ (or equivalently $\beta$) is a function only of the position along the path}; the case $v=\mathrm{constant}$ along the path is a particular sub-case. This amounts to saying  that, in any LCIF along the path, the co-rotating and the counter-rotating beam have the same velocity $v$ in opposite directions. We remark that this assumption naturally generalizes the \textit{equal velocity in opposite directions} condition that was used in \cite{rizzi03b, RRinRRF}, for beams propagating in flat space-time along the rim of a circular interferometer.

If this condition is fulfilled  the coefficient in the second term in (\ref{eq:dtsol1}) is the same for both the co-rotating and the counter-rotating beam, so that this term cancels out and we obtain
\beq
\Delta t = t_{+}-t_{-} = \frac 2 c \oint_{\ell} \frac{g_{0i}dx^{i}}{|g_{00}|}=-\frac 2 c \oint_{\ell} \frac{g_{0i}dx^{i}}{g_{00}} \label{eq:formulafond}
\eeq

Of course the particles take different times for propagating along the path, depending on their speed,  but what we have just shown is that
\textit{ the
difference between these times is always given by eq. (\ref{eq:formulafond}), in any stationary space-time, and for arbitrary paths,} both for matter and light particles, independently of their physical nature. We remark that this result is experimentally well tested (see e.g. \cite{zimmermann65,atwood84,riehle91,hasselbach93,werner79}), and hard to grasp in classical physics: rather, it can be explained in space-time of both special and general relativity, and it is related to the issue of the round-trip synchronization in frames that are not time-orthogonal.  Notice, in fact, that the above condition on the particles speed holds in a LCIF, where clocks are Einstein-synchronized \cite{RRinRRF,RRS}.

Once that the coordinate time difference is known, it is possibile to proceed as in our previous paper \cite{sagnac14}: indeed, eq. (\ref{eq:formulafond}) is the same as eq. (5) of that paper. So we can rephrase the whole discussion, which we summarize as follows.

If the interferometer is located at $P$, on emphasizing the spatial dependence of the metric elements,  the proper time difference that expresses  the \textit{Sagnac time delay} is
\beq
\Delta \tau=-\frac 2 c \sqrt{g_{00}(\mb x_{P})}  \oint_{\ell} \frac{g_{0i}(\mb x)}{ g_{00}(\mb x)}  dx^{i} \label{eq:deltataulocal1}
\eeq
It is possible to write this result in terms  of area enclosed by the path of the beams; to this end, it is useful to define the vector field $\displaystyle \mb h (\mb x) \doteq g_{0i}(\mb x)$, and the scalar field $\displaystyle \varphi(\mb x) \doteq \frac{1}{g_{00}(\mb x)}$, by which we may write the Sagnac time delay in the form
\beq
\Delta \tau=-\frac 2 c \sqrt{\frac{1}{\varphi(\mb x_{P})}}  \oint_{\ell} \varphi \mb h  \cdot d \mb x \label{eq:deltataulocal2}
\eeq
The application of the Stokes theorem and of vector identities allows to write the integral in (\ref{eq:deltataulocal2}) in the form
\beq
\oint_{\ell} \varphi \mb h  \cdot d \mb l = \int_{S} \left[\bm \nabla \wedge (\varphi \mb h) \right] \cdot d \mb S \label{eq:stokes1}
\eeq
where   $\mathbf S$ is the area vector of the surface enclosed by the path of the beams. On using   vector identities  and setting  $\mb b (\mb x)= \bm \nabla \wedge \mb h (\mb x)$, we  eventually obtain
\beq
\Delta \tau=-\frac 2 c \sqrt{\frac{1}{\varphi (\mb x_{P})}} \int _{S} \left[\bm  \nabla \varphi(\mb x) \wedge \mb h  (\mb x)\right] \cdot d \mb S -\frac 2 c \sqrt{\frac{1}{\varphi (\mb x_{P})}}  \int _{S} \left[\varphi (\mb x)  \mb b (\mb x) \right] \cdot d \mb S \label{eq:sagnac2area}
\eeq

The latter expression is the general form of the Sagnac Effect, for both matter and light beams, in terms of surface integrals. We point out that the analogy with the Aharonov-Bohm effect (see \cite{rizzi03a} and references therein), according to which the Sagnac effect can be described in terms of the flux of the  field  {$\mathbf b(\mb x)$} across the interferometer area, is true only  if $\varphi(\mb x)$ is constant over $S$ or its change is negligibly small. Moreover, it is important to emphasize  that,    in the case of the Aharonov-Bohm effect, the {magnetic field} is null along the trajectories of the particles, while in the Sagnac effect the field  {$\mathbf b(\mb x)$ is not null.

Eq. (\ref{eq:sagnac2area}) applies to arbitrary stationary space-times, provided that the above condition ``equal velocity in opposite directions'' is fulfilled. In particular, it can be applied for evaluating the kinematic effects (i.e. neglecting the gravitational ones, which are much smaller, see e.g. \cite{ginger11}) for an interferometer at rest on the Earth surface: the kinematics of the Sagnac Effect is due to the terrestrial diurnal rotation. As we have shown in \cite{sagnac14}, choosing the origin in correspondence of the device (i.e. $\mb{x}_P=0$) and neglecting corrections quadratically depending on the displacements from the origin, we obtain the following expression:
\beq
\Delta \tau=\frac{4}{c^{4}}  \int _{S} \left[{\bm \Omega \left( \bm{\mathcal A} \cdot \mb x \right)} \right] \cdot d \mb S -\frac{4}{c^{2}}  \int _{S} \left[\frac{\bm \Omega }{1+2\frac{\bm{\mathcal A} \cdot \mb x}{c^{2}}} \right] \cdot d \mb S \label{eq:sagnaclocal23}
\eeq
where $\bm{\mathcal{A}}$ is the spatial projection of the device four-acceleration (with respect to the background inertial frame), and $\bm \Omega$ is the generalized rotation rate of the  frame (with respect to a Fermi-Walker transported tetrad). It is then manifest that the Sagnac effect depends, in general, both (i) on the position of the interferometer in the rotating frame through the acceleration $\bm{\mathcal A}$ (whose expression, in this specific case, is related to the laboratory location on the Earth) and (ii) on the interferometer size, since the integrands in (\ref{eq:sagnaclocal23}) are not constant across the interferometer area.

However, a straightforward estimate allows to check that these effects are negligibly small for terrestrial experiments, as we have shown in \cite{sagnac14}: if $R_{\oplus}$ is the terrestrial radius, $\Omega_{\oplus}$ is the terrestrial rotation rate (as measured in an asymptotically flat inertial frame)   we may introduce the dimensionless parameter $\displaystyle  \varepsilon \doteq \frac{1}{c^{2}} \bm{\mathcal A} \cdot \mb x \simeq  \frac{1}{c^{2}}{\Omega_{\oplus}^{2} R_{\oplus}L}{}  \simeq 4 \times 10^{-19} \left(\frac{L}{\mathrm{1 \ m}} \right)$
 where $L$ is the linear size of the interferometer.  Then, the zeroth order in $\varepsilon$ approximation of eq. (\ref{eq:sagnaclocal23}) is
\beq
\Delta \tau = \Delta \tau_{0}= -\frac{4}{c^{2}}  \int _{S} \bm \Omega \cdot d \mb S= \frac{4}{c^{2}}  \int _{S} \bm \Omega_{\oplus} \cdot d \mb S = \frac{4}{c^{2}} \bm \Omega_{\oplus} \cdot \mb S \label{eq:sagnaclocal0}
\eeq
that is the original Sagnac formula (\ref{eq:sagnac2}) as expected, and the analogy with the Aharonov-Bohm effect holds true. Higher order corrections are definitely negligible.

%------------------------Section-------------------------
\section{Conclusions}\label{sec:disconc}
%------------------------Section-------------------------

We focused on the Sagnac effect for matter beams, to evaluate the higher-order corrections of kinematic origin to the basic time delay formula (\ref{eq:sagnac2}), with the aim of generalizing the results that we obtained for light beams in a previous work. To this end, we showed that the difference between the propagation times of two matter  beams moving in opposite directions along a closed spatial path is the same, independently of the nature of the beams, provided that (i) the space-time metric in the laboratory is stationary and (ii) the speed of the beams in the two opposite directions (as measured in a LCIF)  is the same at any position along the path. In particular, this time difference is the same both for matter and light beams, and it is simply related to the time gap arising in non time-orthogonal frames. Actually, we  focused on the kinematic aspects: in other words we considered propagation times of particles representative of the interfering beams. For light beams, dynamic aspects are properly studied by solving the Maxwell equations (see e.g. \cite{chow85}); as for matter beams, quantum mechanical wave equations are needed (see e.g. \cite{stodolsky}). It is worthwhile mentioning that, in order to properly speak of interference, we must  obtain a phase difference: in other words, a frequency is needed: we can naturally attribute a frequency to light beams, while for matter beams  we have to consider the de Broglie waves frequencies, even though some subtleties must be taken into account (see e.g. \cite{scorgie,scorgiebis}).

On the basis of our results, we wrote an exact expression of the Sagnac effect in terms of surface integrals across the interferometer area, valid in arbitrary stationary space-time, which enabled us to investigate the role of the  position and size of the interferometer and to discuss the analogy with the Aharonov-Bohm effect. Of course, the same conclusions of our previous paper \cite{sagnac14} apply:  (i) in general, the Sagnac effect is influenced by both the position of the interferometer in the rotating frame and its extent; (ii)  the analogy with Aharonov-Bohm effect holds true to lowest approximation order only. However, in actual experimental situations, the higher order corrections are negligible and the effect is safely described by the expression (\ref{eq:sagnac2}), both for matter and light beams. In this approximation the analogy with the Aharonov-Bohm effect can be applied, even though the two effects are quite different in general.

\ed

%------------------------Section-------------------------
\section{Conclusions}\label{sec:conc}
%------------------------Section-------------------------

We studied the Sagnac effect for light beams in flat space-time, in order to evaluate the possible higher order corrections of kinematic origin to the Sagnac formula (\ref{eq:sagnac2}). In particular, we focused on the relevance of these terms for terrestrial experiments that are now being planned. To this end, we worked out the necessary formalism in the context of the local laboratory frame and, in order to make a connection with the Aharonov-Bohm effect, we derived an exact expression of the Sagnac effect in terms of surface integrals, across the interferometer area. We showed that the analogy with the Aharonov-Bohm effect holds true to lowest approximation order only and that,  in general, the Sagnac effect is influenced by both  the position of the interferometer in the rotating frame and  its extent. The expressions that we obtained can be developed in power series of a suitable dimensionless parameter, to obtain all kinematic corrections. However, as for the accuracy available for the experiments that are now under investigation, the lowest order approximation is sufficient, so that the special relativistic or kinematic Sagnac effect is consistently described by the expression (\ref{eq:sagnac2}).

\end{document}